# Further developments and tests of microstrip gas counters with resistive electrodes

R. Oliveira,[1] V. Peskov,[1,2] Pietropaolo,[3] P.Picchi[4]
[1]CERN, Geneva, Switzerland
[2]UNAM, Mexico
[3]INFN Padova, Padova, Italy
[4]INFN Frascati, Frascati, Italy

**Absract**

We present results from further tests of Microstrip Gas Counters (MSGCs) with resistive electrodes. The maim advantage of this detector is that it is spark-protected: in contrast to "classical" MSGCs with metallic electrodes, sparks in this new detector do not destroy its electrodes. As a consequence the MSGC with resistive electrodes is more reliable in operation which may open new avenues in applications. One of them which is under investigation now is the use of Resistive electrodes MSGC (R-MSGC) as photodetector in some particular designs of noble liquid dark matter detectors.

## I. Introduction

MSGC was the first micropattern gaseous detector developed some time ago by A. Oed [1]. His pioneering idea to apply microelectronic technology for the manufacture of gaseous detectors triggered a great interest in this direction and many other types of micropettern detectors based on this approach were later developed, for example MICROMEGAS, GEM and so on (see recent review papers [2,3]).

A lot of effort from different groups was invested in understanding what limits the maximum achievable gain of micropattern detectors in general and in MSGS in particular. Typically it was possible to achieve a maximum gain of $10^4$ with MSGCs irradiated with 6 keV photons; at higher gains sparks appear which usually strongly damage or even fully destroy the detector (see Fig. 1).
Studies show (see for example [4]) that the maximum achievable gain of micropattern gaseous detectors is governed by the so-called Raether limit. In the case of the MSGCs, however, an additional mechanism can strongly contribute to breakdown developments – surface streamers [5-7]. The surface streamers appear at voltages applied between the anode and the cathode strips $V>V_{crit}$, where $V_{crit}$ depends on gas surface geometry and surface property/quality: its material, cleanness, roughness and so on. Favorable conditions for the development of surface streamers in MSGC are a low ratio of anode strip width to the strip's pitch and poor surface quality. Due to the surface streamers the maximum achievable gain of the MSGC can be less than that expected from the Raether limit.
The main conclusion from these studies is that the maximum achievable gain of small pitch MSGC(<0.5mm) due to the physics of discharges, cannot be more that $10^4$ for 6 keV photons. Thus efforts should be focused on making MSGS robust enough so that they can withstand the sparks.

There has been quite a lot of work in this direction. For example, Bellazini et al [8] have developed MSGCs, the cathode strips of which were covered with thin dielectric discharge protective films. Although the voltage range over which a chamber may be safely operated was extended by more than 100V [9] this method had a limited success. The best practical solution found so far is to combine MSGC with a gas avalanche preamplification structure, which could be GEM [10-12] or just two parallel meshes (parallel-plate avalanche chamber operating at some gain) [13]. Such combination of two detectors operating in cascade mode has two advantages :1) it allows MSGC sto operate at less gain, hence at voltages $V<V_{crit}$, 2) it increases the Rather limit of the MSGCs due to the diffusion of avalanche electrons exiting the preamlification structure and moving towards the MSGC [13]. MSGCs combined with GEMs were successfully implemented in the HERRA-B inner tracker [12] and this cascaded detector demonstrated good and reliable performance.

One should note that as a matter of fact other micropattern gaseous detectors, which were developed after the MSGC invention , typically have a maximum achievable gain comparable to the MSGC one (~$10^4$ for 6 keV photons) and it also governed the Raether limit [4]: at gains exceeding the Raether limit frequent sparks are unavoidable. This is why in the last few years considerable interest has arisen in micropattern gaseous detectors with resistive electrodes. This modification does not increase the maximum achievable gain (which is still governed by the Rather limit), but allows the detector to withstand sparks without being destroyed. We havealready developed and tested prototypes of GEM and MICROMEGAS detectors with resistive electrodes and demonstrated that in these detectors the sparks current was reduced, depending on the particular design, 100-1000 times. Following this line we also recently tried to develop MSGCs with resistive electrodes. We named this detector Resistive MSGC or R-MSCGS. Our first prototype manufactured by the screen printing technique is described in [14].The maximum achievable gas gain of this detector was only about 10, however it withstood any amount of sparks even if they were created in air atmosphere and were clearly seen by the eye. The studies show that the gain limitation came from surface streamers due to the bad surface quality of these first prototypes. In this report we describe results obtained with better quality R-MSGC allowing gains up to $10^3$ to be achieved.

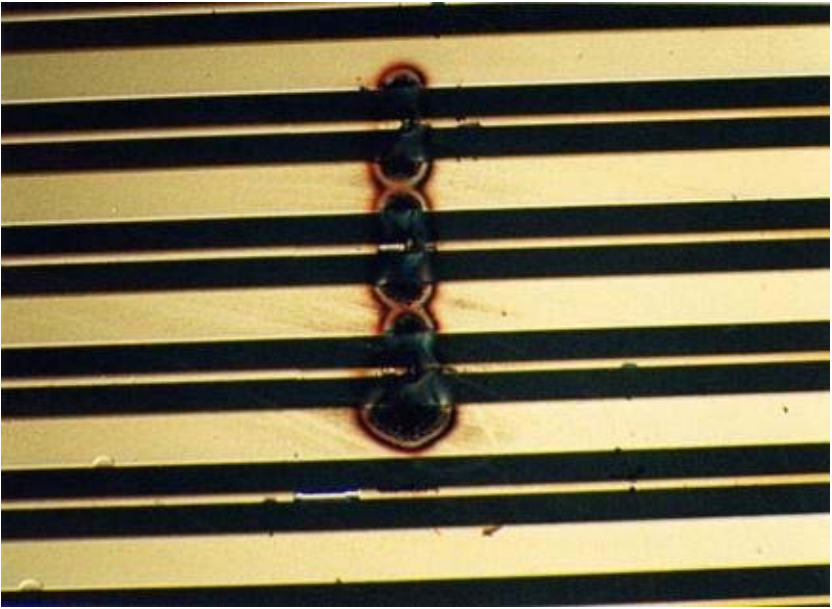

Fig. 1. A photograph of MSGC damaged by discharges (courtesy of L. Ropelewski)

These spark protective R-MSGCs can be very attractive for some applications. For example, in the last paragraph we will briefly describe their possible application for some designs of cryogenic detectors.

**II. Materials and methods**

R-MSGCs were manufactured on the top of usual 1.6 mm thick fiber glass plates (FR-4) using a combination photolithography and screen printing technique.
Two types of detectors were manufactured and extensively tested. The first one had both the anode and the cathode strips made of resistive material (paste ESL RS 12116, surface resistivity 1MΩ/□), whereas the second detector had metallic anode strips and resistive cathode strips
 The manufacturing steps of the first detector are shown schematically in Fig. 2. First Cu strips 50 μm wide were created on the top of the fiber glass plate by the photolithographic method (see Fig 2a and 2b). Then in contact with the side surfaces of each Cu strip, dielectric layers (Pyralux PC1025 Photoimageable coverlay by DuPont) were manufactured; they have a thickness of 175μm and a height of 70 μm. Finally the detector surface was coated with resistive paste and polished so that the anode and the cathode resistive strips become separated by the Coverlay dielectric layer. The width of the anode strip was 80 μm, the width of the cathode strip was 300 μm and the pitch was 600 μm.

The schematic drawing of the top view of this detector is shown in Fig.3. In contrast to "classical" MSGCs no special care was taken with the design of the end of anode strips: they simply had rounded shape ends (note that in "classical MSGCs each end of the anode strip is shaped in the form of circle and the cathode strips have a co-centric round shape with a larger radius).
To ensure a proper reduction of the energy of the spark which may happen at the electrically weak spots (one end of the each strip) the resistivity of the region schematically marked as "A" in Fig.3 of the anode strip and the zone C of the cathode strip (Fig. 3) were in the range 4-6 MΩ.[1]

The schematic presentation of steps in manufacturing the second detector is presented in Fig. 4. First by photolithographic technology Cu strips 200 μm in width were created on the top of the FR-4 plate (Figs 4a and b). Then the gaps between the strips were filled with the resistive paste (Fig. 4c). As a next step the middle parts of the Cu strips were coated with 50 μm width layer of photoresist (Fig. 4d) and the rest of the area of the Cu strip was etched (Fig 4e); in such a way metallic anode strips 50 μm in width were created. Finally the gaps between Cu anode strips and the cathode resistive strips were filled with glue FR-4 and after it hardened the entire surface was mechanically polished. As will be shown later this was the more successful design than the detector #1 so far allowing higher gas gains to be reached than with detector #1.
A photograph of the detector #2 is shown in Fig. 5 and in Figs. 6-9 are presented magnified photographs of various regions of this detector: the central one (Fig. 6), the border of regions B and C (Fig. 7) and regions B (Fig.8) and C (Fig. 9).

---

[1] This is especially profitable when only one of electrodes is resistive and the another one is metallic (see the text below)

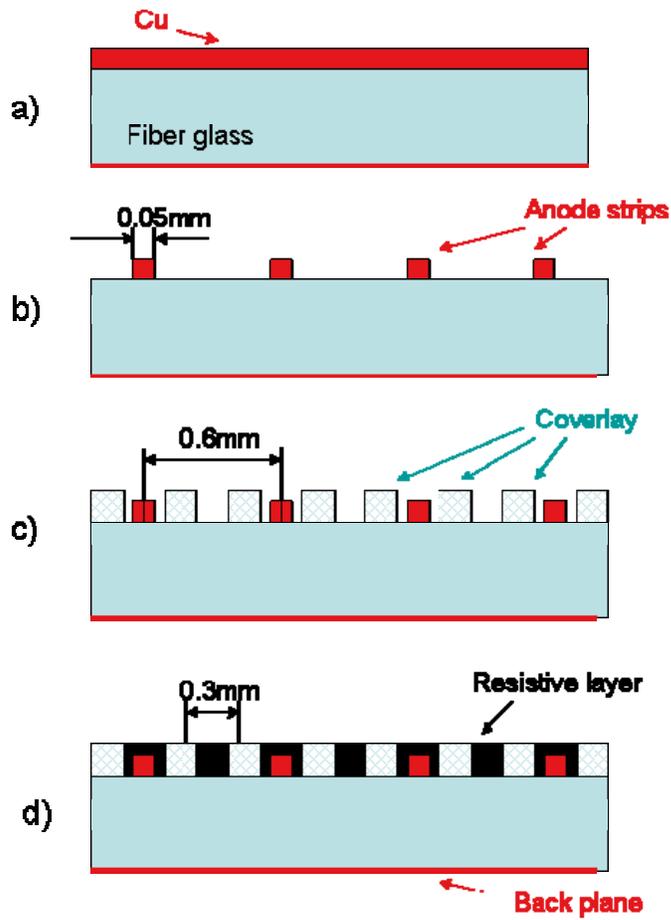

Fig. 2. Schematic illustration of the main manufacturing steps of the R-MSGC#1 with resistive anode and cathode strips

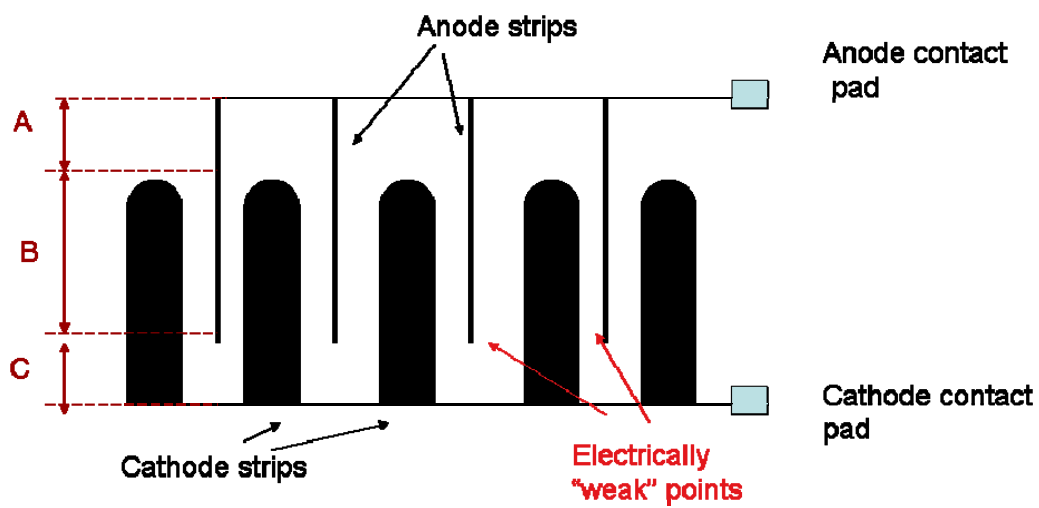

Fig. 3. A top view of the R-MSGC with the anode and the cathode resistive strips

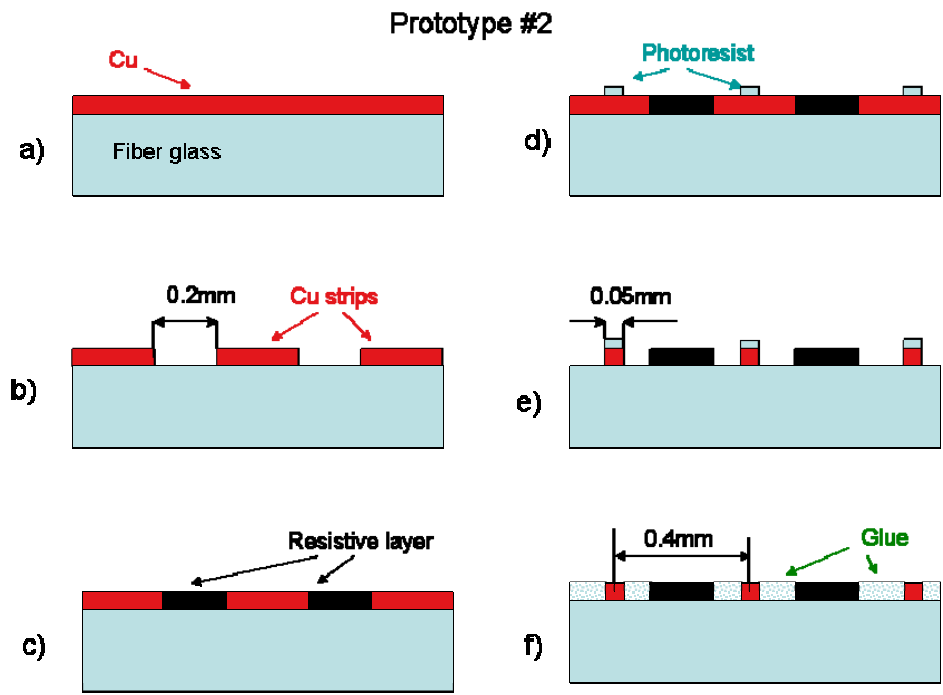

Fig. 4. Main manufacturing steps of the R-MSGC#2 in which the gap between the anode and the cathode strips was filled with glue RF-4 and then after hardening the glue the top surface was polished

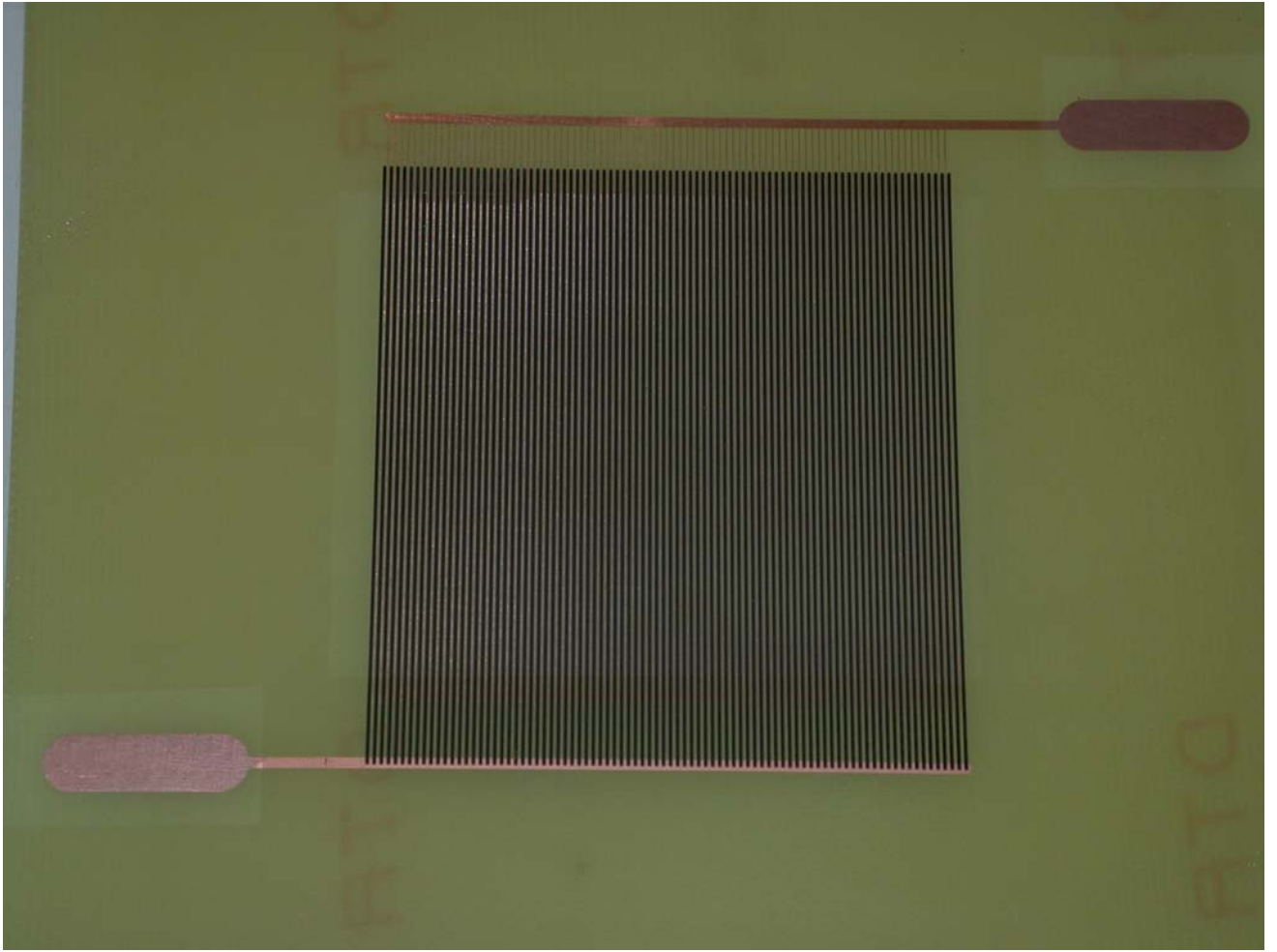

Fig.5. A photograph of R-MSGC #2 with Cu anode strip (50μm width, 400 μm pitch) and resistive cathode strips ( 200 μm width ).

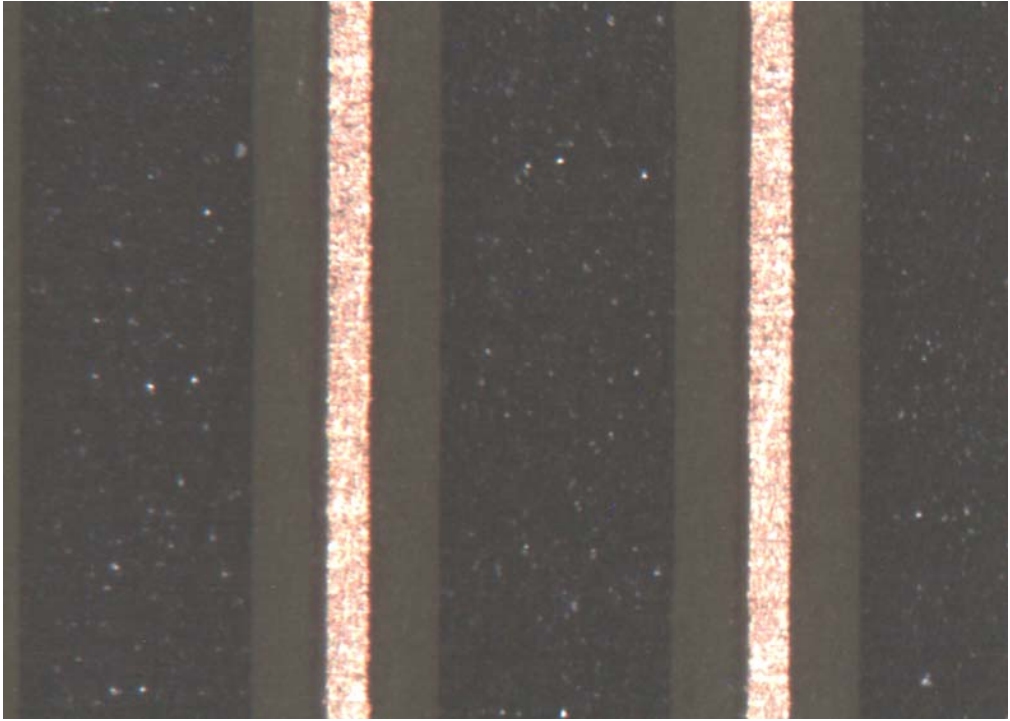

Fig..6. A magnified photograph of the central region of R-MSGC #2 with Cu anode strips (a narrow strip) and resistive cathode strips (wide black strips).

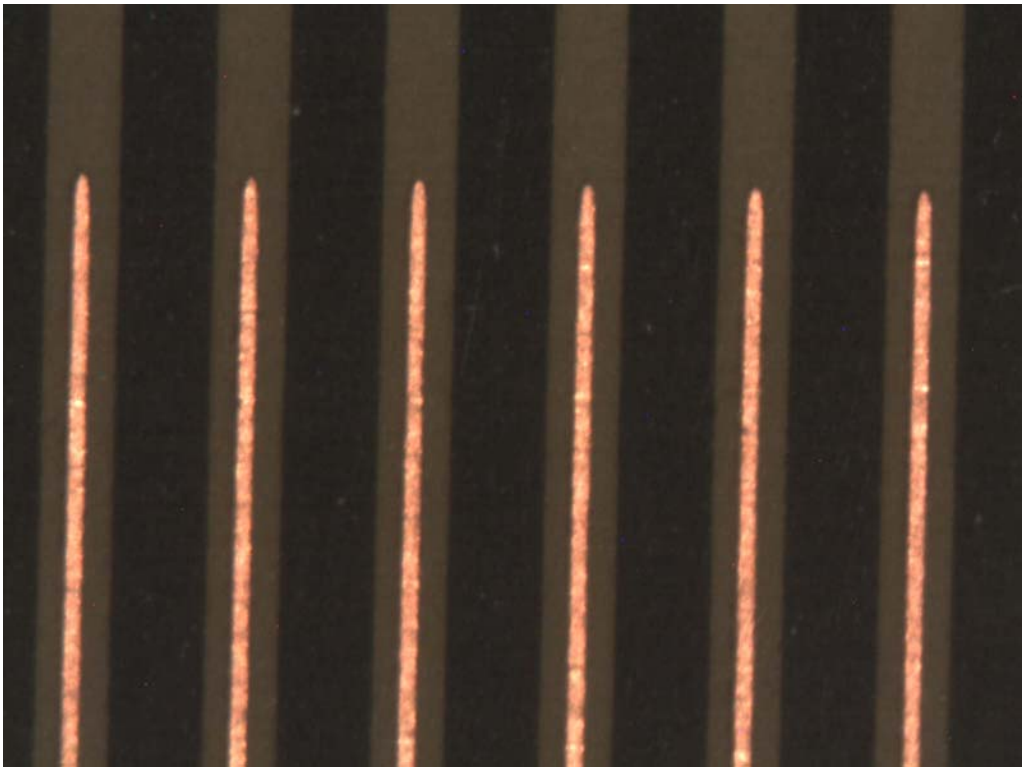

Fig.7. A photograph of the border of regions B and C (see Fig.3).

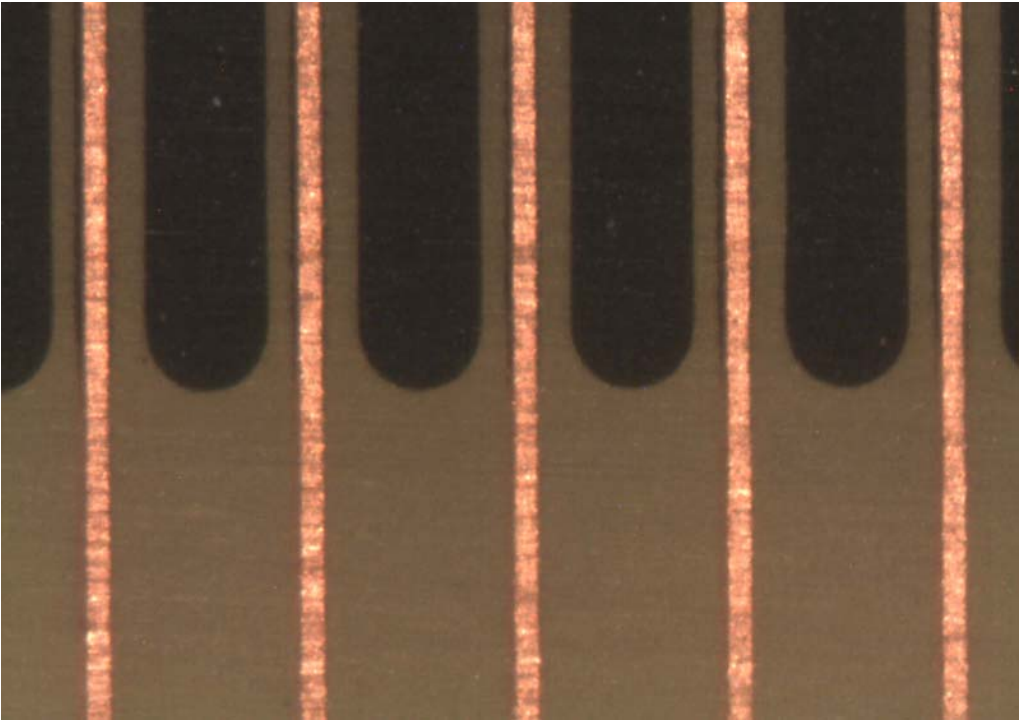

Fig. 8. Photo of region B

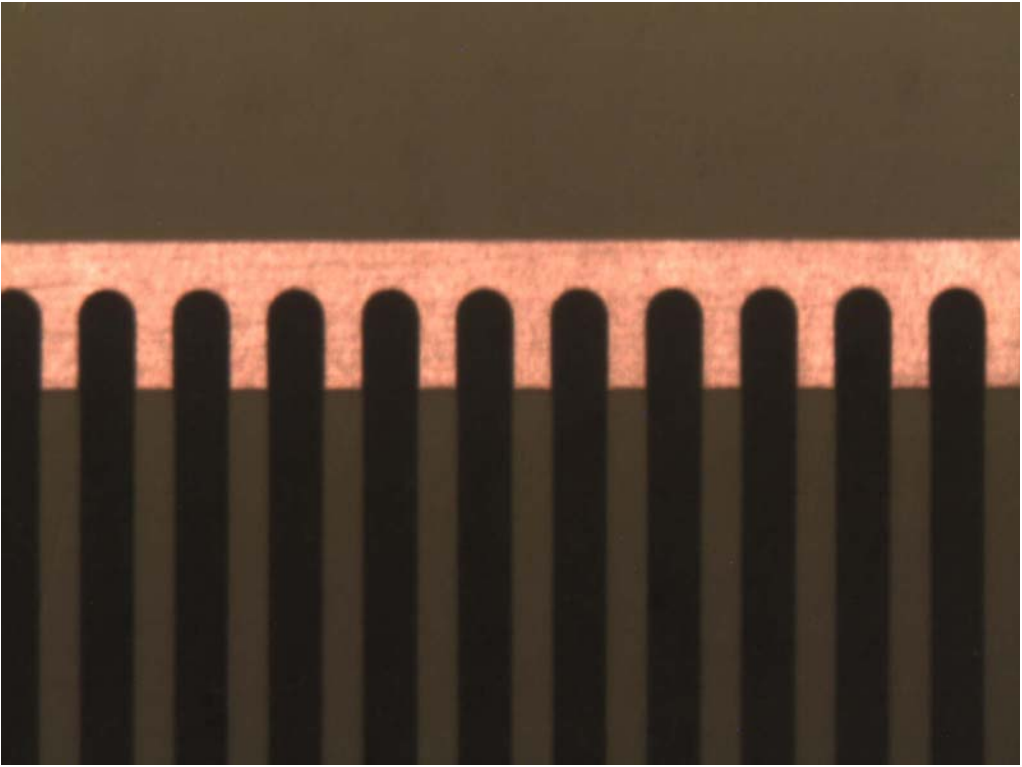

Fig.9. Magnified photo of the cathode strip at the end of the region C, where these resistive strips are connected via the Cu strip for the HV supply

In both designs all cathode strips were connected together and to the HV supply and all anode strips also were together connected to the charge sensitive amplifier.
The R-MSGCs were tested in the gas chamber described in [15]. The small modification was that the CaF$_2$ window was replaced by a beryllium window combined with a lead collimator. Measurements were performed in Ne and mixtures of Ne with CH$_4$. The primary ionization in the detector volume was produced by $^{241}$Am (alpha particles) and by $^{55}$Fe source (6 keVphotons). In studies of rate characteristics of the R-MSGC an X-ray gun was used (XTF5011, Oxford Instruments) having a Cu anode.

We also tested the operation of R-MSGC combined with a parallel-plate preamplifaction structures. The latter was manufactured from two resistive meshes [15,16] with a gap between them of either 1 or 0.7mm.

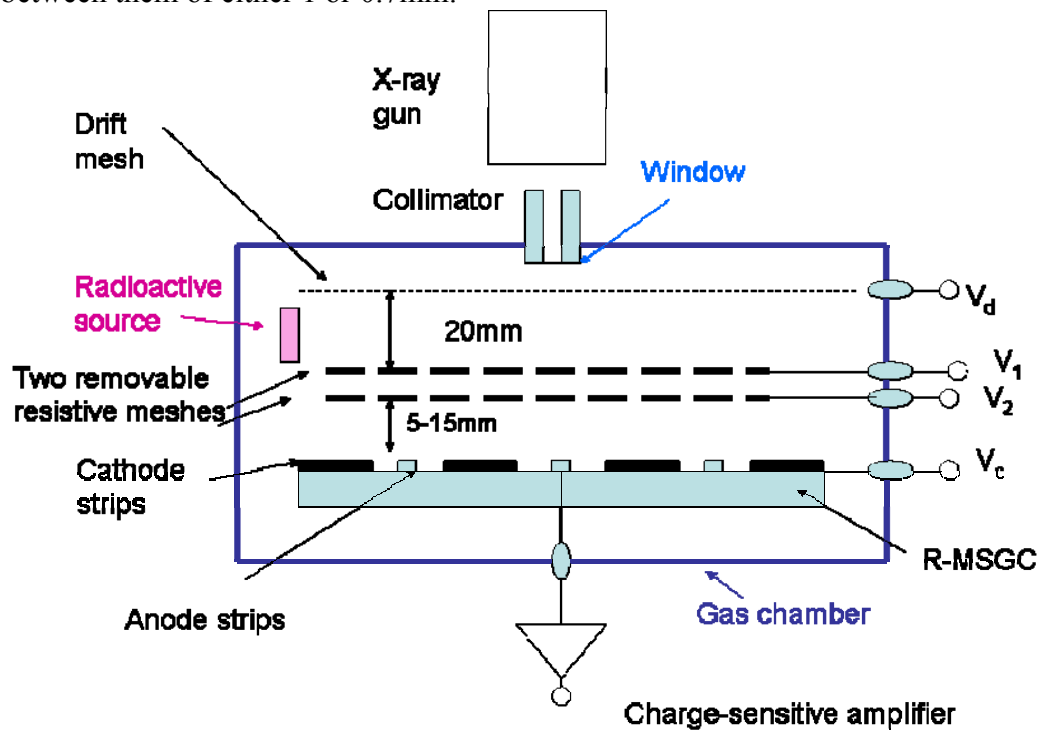

Fig.10. A schematic drawing of the experimental setup used in studies of R-MSGCs

### III. Preliminary results

Fig. 11 shows the main results obtained with R-MSGC #1 having both the anode and the cathode resistive strips. Filled rhombuses and squares represent the gain measured in Ne and Ne+6% CH$_4$ respectively. The highest points in these curves correspond to the gains at which first discharges appear (at a rate ~ 1 per 10min). Curves with open circles show the surface streamer rate: the discharges in R-MSGC appear when the streamer rate becomes ≥~1per min. As can be seen, gains around 10 were achieved with this design. Obviously these maximum achievable gas gains are not high and we believe this is due to the poor surface quality. However in the case of sparks their energy, due to electrode resistivity, was strongly reduced so even after a long-term continuous sparking the detector was never destroyed.

In an attempt to diminish the surface streamer rate the R-MSGC was, in additional, very carefully cleaned with deionized water and dried in the oven. Results obtained after additional cleaning are shown in Fig 12. As can be seen, after cleaning the maximum achievable gain was increased 3-4- times.

As was mentioned in the introduction, in the past in order to boost the gains of ordinary MSGCs (which was usually $\leq 10^4$ for 6 keV photons) different authors tried to use various preamplication structures [10-13]: for example, GEM or a PPAC constructed from two metallic meshes. However, this approach does not guarantee full spark protection: at high overall gains sparks still may occur in the MSGC itself or propagate from the preamplification structure to the MSGCs. Some studies related to this effect are summarized in preprint [4].
A PPAC, recently developed by us, and assembled from resistive meshes ( we called them resistive mesh PPAC or RM-PPAC) offers a real spark protection: with this preamplification structure no spark propagation was observed [15,17].
The open rhombuses in Figs. 11 and 12 show the gain of a R-MSGC combined with a RM-PPAC; as can be seen this allowed the maximum achievable gains to be increased by a factor of 100. There are two main reason for this: 1) the Raether limit is usually higher in cascaded detectors than in single ones (due to the diffusion of avalanche electrons) [13], 2) the voltage on the R-MSGC electrodes can be kept below $V_{crit}$ which reduces the surface streamer rate and thus the probability of discharges.

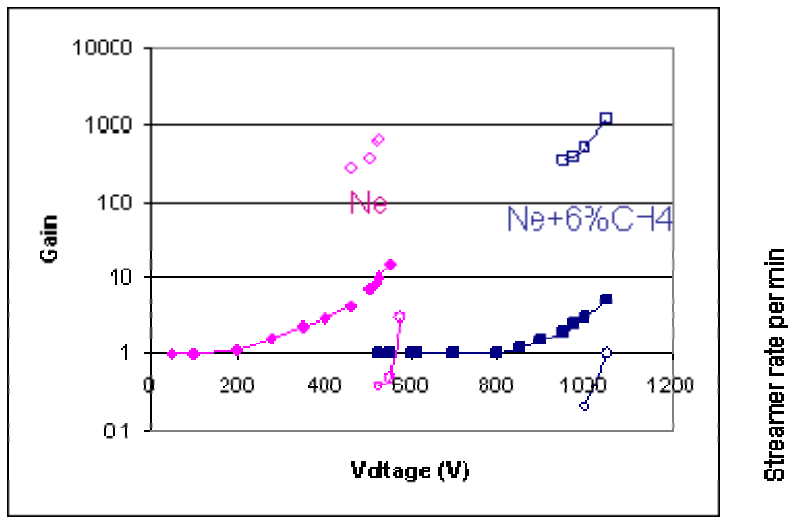

Fig.11 . Gain vs. voltage curves measured with R-MSGC#1 having the anode and the cathode resistive strips in Ne (filled rose rhombus) and in Ne+6%CH$_4$ (filled blue squares). In the same figure open circles represent surface streamers rate and open rhombus and squares- gains achieved when the R-MSGC was combined with the RM-PPAC.

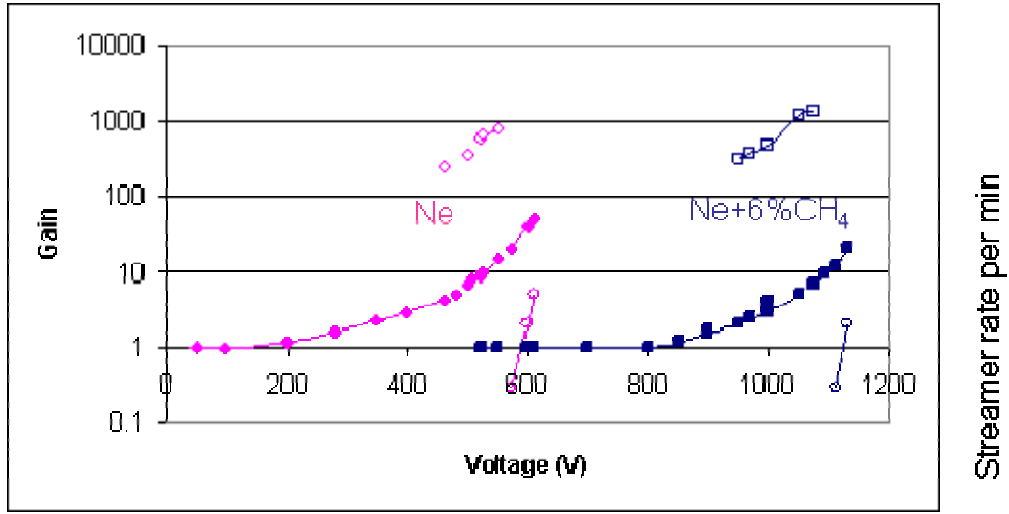

Fig 12. Gain vs. voltage curves measured with R-MSGC#1 after additional cleaning of its surface. All symbols have the same meaning as in Fig.11.

We also tested operation of R-MSGCs when a such a high voltage was applied to the drift electrode that some amplification occurred in the drift region which then began to seve serving as a preamplification structure-see Fig.13. This mode of operation in the case of ordinary MSGCs was earlier studied by Bellazzini [ 19] and is very attractive for some applications when the vertex of tracks of charged particles (or track of photoelectrons in the case of the detection of collimated X-rays) )and the drift plane is necessary to detect with high position resolution. The examples of such applications can be found in references [20-23].

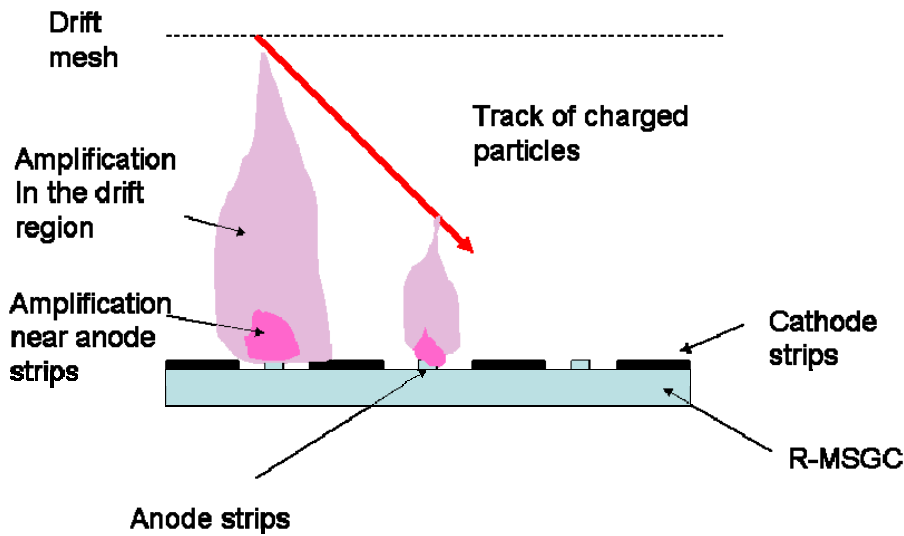

Fig.13.A schematic drawing of R-MSGC operating with preamplification in the drift region

The reason why the vertex of the charged particles and the drift plane can be detected with high position resolution is simple: in the case of the parallel-plate geometry the multiplication factor depends exponentially on the distance of the primary electrons from the cathode. As a result, the main contribution to the signal is from the primary electrons created near the cathode; the other part of the track, even if it is very long but inclined (and most of the tracks- especially photoelectron tracks- are inclined), contributes very little to the signal amplitude. Thus in this particular geometry one mostly detects the vertex of the photoelectron track.

Some of our results obtained when R-MSGC operated with preamplification in the drift region are shown in Fig 14. As can be seen, overall gains up to $10^3$ were achieved in this mode. At a higher gain discharges appeared, but they never damaged the R-MSGC[2].

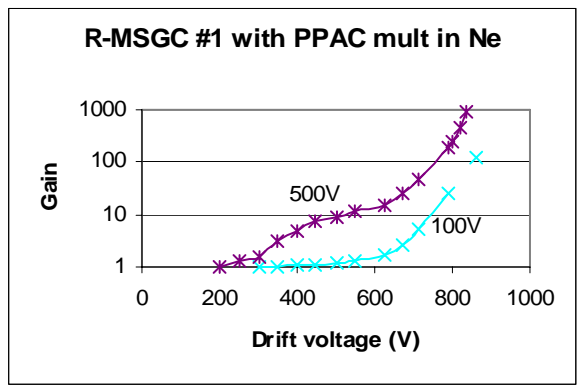

Fig. 14. Gain curves for the R-MSGC#1 operated in Ne in PPAC mode. The number near the curves indicate the voltage applied between the R-MSGC strips

Considerably higher gains were achieved with the R-MSGC#2 having resistive cathode strips and Cu anode strips(see Figs. 15 and 16); this was mainly because the anode strips had narrower width and in this case surface streamers appear at higher gains[6,18].

---

[2] This indicated that resistive strips can be used inside the avalanche gag for the readout of PPACs

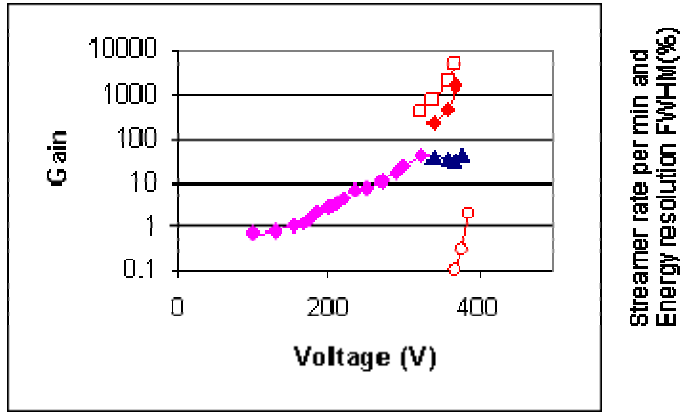

Fig.15. Some results obtained with R-MSGC#2 operating in Ne: filled rose rhombus –gain vs. voltage measured with alpha particles emitted from $^{241}$Am, red rhombus-gain measured with 6 keV photons from $^{55}$Fe. Open circles-streamers rate, blue triangles represent the measurements of the energy resolution: FWHM of the $^{55}$Fe pulse height.
Open rhombuses-overall gains achieved with RM-PPAC (1mm gap) preamplification structure.

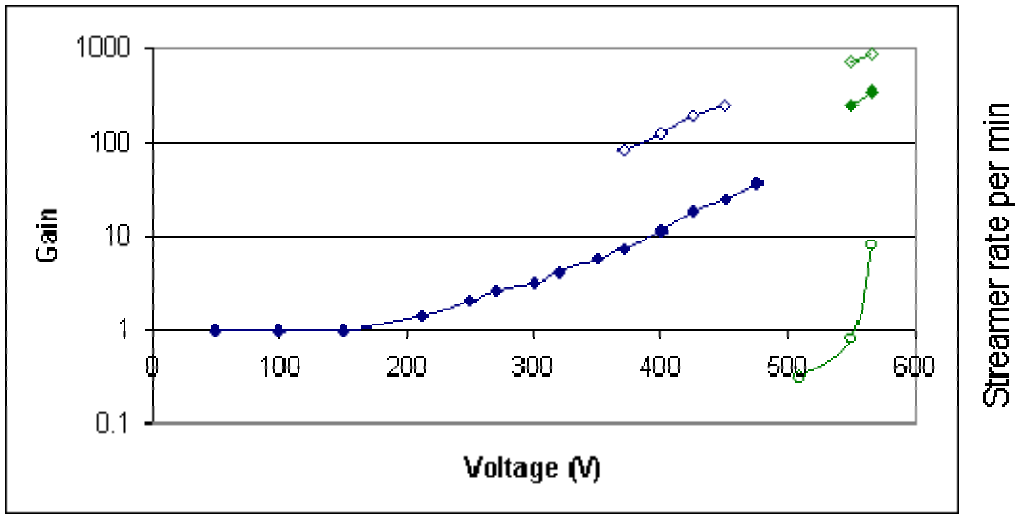

Fig.16. Gain vs. voltage curves and streamer rates measured with R-MSGC#2 operating in Ne+5%CH$_4$. Filed blue rhombus –gain vs. voltage measured with alpha particles emitted, green rhombus-gain measured with 6keV photons, open blue and green rhombuses-overall gains achieved with RM-PPAC (0.7 mm gap) preamplification structure. Open green circles-surface streamer rate.

Some results of measurements of the rate characteristic of the R-MSGC#2 are shown in Fig.17. Practically no drop of pulse amplitude was observed up to a counting rate of $10^4$Hz/cm$^2$- the maximum available rate from the Cu tube in the setup used by us.

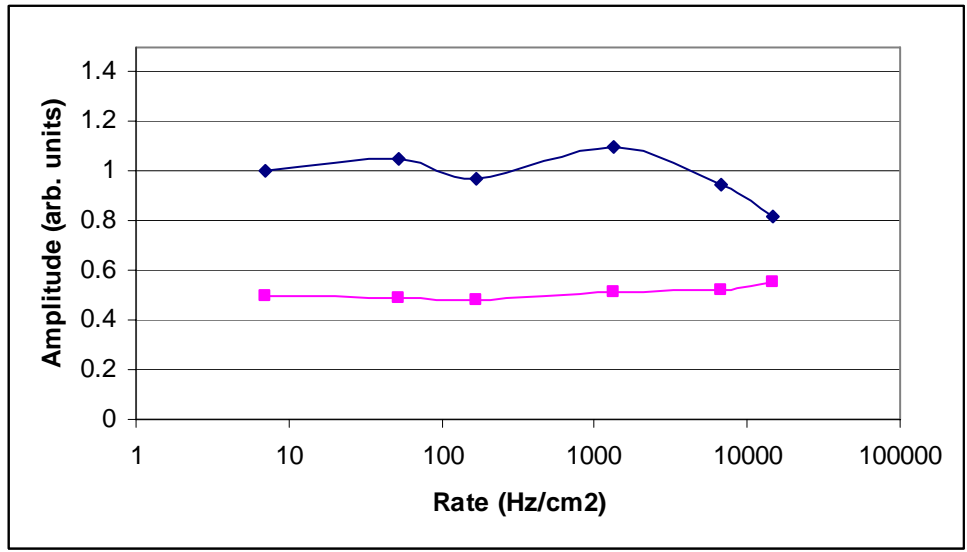

Fig. 17. Mean pulse amplitude vs. the counting rate measured in Ne (blue rhombus) and in Ne+5%$CH_4$ (rose rhombuses).

To demonstrate the spark protection capability of the R-MSGC#2, we intensively operated it in spark mode (as was done earlier with R-MSGC#1). No destruction of electrodes was observed even after several hundred sparks.

## IV. Discussion

Preliminary tests described above demonstrate the feasibility of building spark protective MSGCs. In these detectors due to the resistivity of electrodes and the small capacity between the strips the spark energy was strongly reduced so the strips were not damaged even after a few hundred sparks.
We are planning to apply this technology to the manufacture of a CONBRA-type [24] detectors with resistive electrodes The spark-protected COBRA will be an attractive option for the detection of charge and light from the LXe TPC with a CsI photocathode immersed inside the liquid. The concept of this detector is described in [25] (see Fig. 18) and the operation of CsI photocathode inside noble liquids already demonstrated in several works [26,27]. In such TPCs, combined with the CsI photocathode, it is very difficult to use conventional parallel-plate structures or GEMs for the primary charge and light detection since the scintillation light from these structures will cause a photoeffect from the CsI photocathode and a feedback loop. In CONBRA- type detectors, the amplification region will be geometrically shielded from the CsI photocathode (see Fig.19) and accordingly the feedback will be reduced[3]. Since such a detector should operate in large dynamic ranges (detection from single electrons up to hundreds or thousands of primary electrons) the discharges might be unavoidable ( due to the so called the Rather limit -see [28] for details) so it will be essential to have spark protected detectors.
Another option for the LXe TPC which is currently under study in our group is to use LXe doped with low ionization potential substances. In this detector feedback will also be a problem and thus COBRAD with resistive electrodes will be an attractive option.

---
[3] In additional one can used HV gating in the in the structure placed in front of COBRA (or in COBRA itself) to disable the detector for some time after the detecting the primary and secondary scintillation light

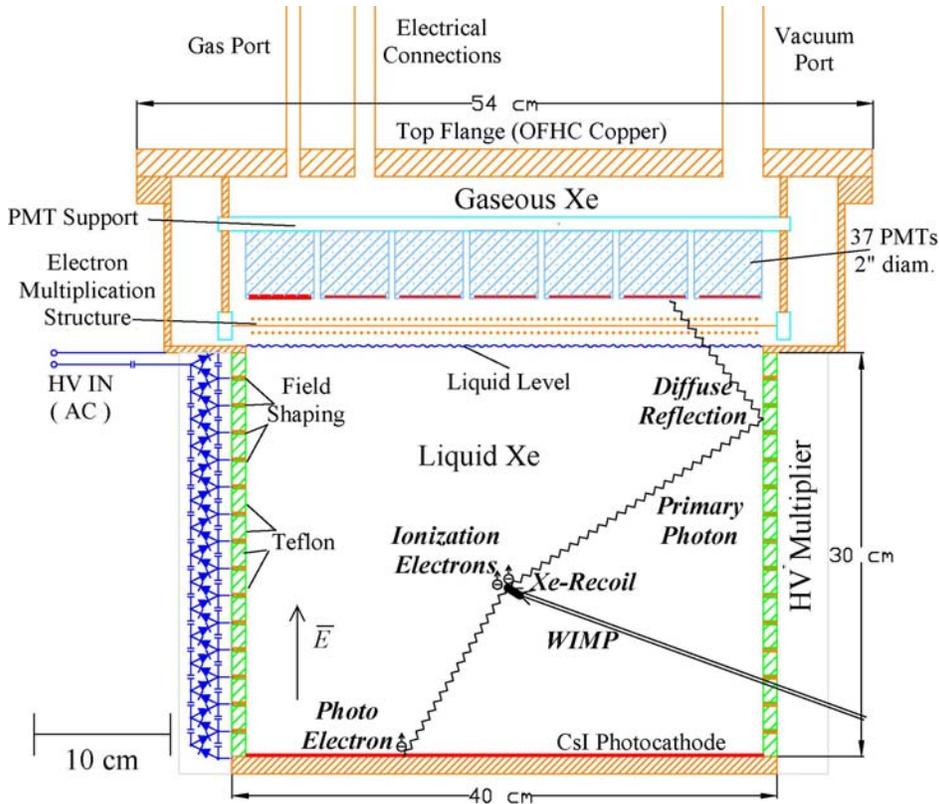

Fig.18. A schematic drawing of the LXe WIMP detector with a CsI photocathode immersed inside the liquid for the detection of primary scintillation light produce by various interactions inside the TPC (from [25]).

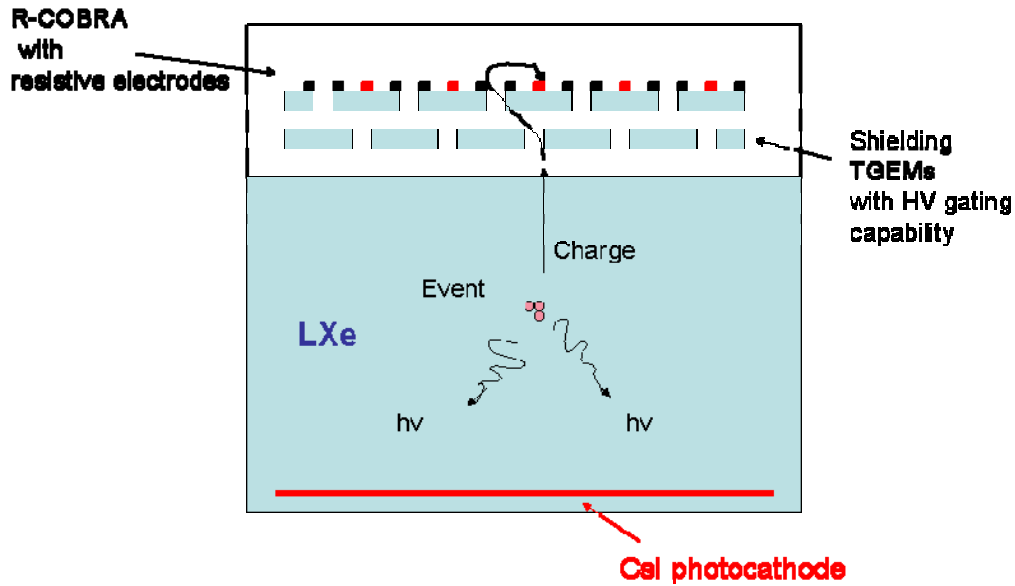

Fig.19. Another possible version of the LXe detector (under study) in which the primary charge and the primary scintillation light form the LXe is detected by the resistive COBRA detector